\documentstyle[amssymb,aps,12pt]{revtex}

\begin{document}
\title{Occupation numbers in Self Consistent RPA}
\author{J. Dukelsky}
\address{Instituto de Estructura de la Materia, Consejo Superior de\\
Investigaciones Cient\'\i ficas,\\
Serrano 123, 28006 Madrid, Spain}
\author{J. G. Hirsch}
\address{Instituto de Ciencias Nucleares, Universidad Nacional\\
Aut\'onoma de M\'exico,\\
Apdo. Postal 70-543 M\'{e}xico 04510 D.F.}
\author{P. Schuck }
\address{Institut des Sciences Nucl\'{e}aires, IN2P3-CNRS,\\
53 Av. des Martyrs, 38026 Grenoble-Cedex, France}
\maketitle

\bigskip

\begin{abstract}
A method is proposed which allows to calculate within the SCRPA theory the
occupation numbers via the single particle Green function. This scheme
complies with the Hugenholtz van Hove theorem. In an application to the
Lipkin model it is found that this prescription gives consistently better
results than two other commonly used approximations: lowest order boson
expansion and the number operator method.

\end{abstract}



\section{Introduction}

The solution of the many body problem beyond the meanfield level is not a
very well settled problem. Though the meanfield approach for all kinds of
many body problems is quite uniquely defined, the determination of the
higher order correlation functions is not. Besides the usual partial
resummation of Feynman graphs ( e.g. ring summation in RPA) there also exist
variational ans\"atze such as those introduced by  Jastrow, Gutzwiller, 
together with the Resonating Valence Bond approach, etc. \cite{Blai86}.
However only in
rare cases these variational approaches can be worked through to the end by
minimizing the groundstate energy so that any new route can have interesting
perspectives. In most cases there remains the additional problem of how to
determine the excited states. One of the attractive features of  the
Raleigh-Ritz variational Hartree-Fock
(HF) theory is indeed that it yields, consistently within the same theory,
groundstate and excited states (quasiparticle excitations).

Since some time we have elaborated on a theory for two body correlations
functions which in a certain sense can be considered as an extension of HF
theory to two body clusters. We for instance obtain selfconsistent
nonlinear equations for the correlation functions which simultaneously
determine the correlated groundstate energy and the spectrum of excitations.
We named this approach Self - Consistent Random Phase Approximation (SCRPA),
since it is a consistent generalization of the standard linear RPA
approach \cite{Schu73,Duk90,Duk96,Duk98a,Duk98b}.
This formalism was also developed independently by a second group of authors
which coined for it the name Cluster Hartree-Fock (CHF) which seems also
very appropriate \cite{Rop95}. 
This type of theory took its roots several decades back starting with the work 
of Hara \cite{Har64}. Considerable progress was achieved by D. Rowe using the 
equation of motion method which is summarized in \cite{Row68a}. Some years 
later the theory was rederived using the method of many body Green functions
\cite{Schu71,Schu73}. Since that time not much progress was made on the formal
aspect of the theory until the more recent works cited above. 

The SCRPA has 
lately
given a series of interesting results for various many-body problems
\cite{Kru94,Duk96,Duk98a}. 
Nevertheless some
open problems persisted in the past with this formalism concerning for
instance the consistent evaluation of single particle quantities such as the
single particle density matrix or the occupation numbers. An approximation
which lately came very much in use in relating these quantities back to
SCRPA (or to its poorer but numerically easier variant the so called
Renormalized RPA (RRPA) \cite{Har64}) is based on the particle number
method which long time ago already was advocated by D. J.Rowe
\cite{Row68}. Very recently we have proposed
and applied a different method which calculates these quantities via the
single particle Green's function with a mass operator coupling back to the
SCRPA \cite{Duk96,Duk98b}. In those works, however, neither a detailed
derivation nor an assessment of its quality was given . On the other hand it
has been pointed out that certain consistency relations are indeed
fulfilled.

The purpose of the present paper is therefore to give a quite detailed
derivation and to make a systematic investigation in a model case of the
Green's function approach and to contrast it with other methods.

The paper is organized as follows: In section II the SCRPA equations are
deduced, their coupling with the single particle Green's functions is
presented in section III, the application to the Lipkin model is developed
in section IV, the numerical results in section V and the conclusions
are given in section VI.

\section{Outline of the problem}

Self Consistent RPA can be derived in various ways. The method which
probably exhibits most clearly the analogy with ordinary HF theory is the
one due to Baranger \cite{Bar70}. Let us first rederive in this way
single particle HF.
To this end we define a mean single particle energy in the following way

\begin{equation}
\epsilon _{\mu }=\frac{\sum\limits_{\nu ,k}\left\{ \left( E_{\nu
}^{N+1}-E_{0}^{N}\right) \left| \left\langle N,0\right| \varphi _{k}^{\mu
}a_{k}\left| N+1,\nu \right\rangle \right| ^{2}+\left( E_{0}^{N}-E_{\nu
}^{N-1}\right) \left| \left\langle N,0\right| \varphi _{k}^{\mu
*}a_{k}^{\dagger }\left| N-1,\nu \right\rangle \right| ^{2}\right\} }{%
\sum\limits_{\nu ,k}\left\{ \left| \left\langle N,0\right| \varphi
_{k}^{\mu
}a_{k}\left| N+1,\nu \right\rangle \right| ^{2}+\left| \left\langle
N,0\right| \varphi _{k}^{\mu *}a_{k}^{\dagger }\left| N-1,\nu \right\rangle
\right| ^{2}\right\} }  \label{e}
\end{equation}
where $E_{\nu }^{N}$ and $\left| N,\nu \right\rangle $ are in principle
exact eigenenergies and eigenstates of the Hamiltonian for a system with
N particles. For the groundstate
we have $\nu =0$ and $a_{k}^{\dagger }$ is a single particle creation
operator. Minimizing (\ref{e}) with respect to the amplitudes $\varphi
_{k}^{\mu }$ and $\varphi _{k}^{\mu *}$ leads directly to the following
eigenvalue problem

\begin{equation}
\sum_{k^{\prime }}\left\langle 0\right| \left\{ a_{k},\left[ H,a_{k^{\prime
}}^{\dagger }\right] \right\} \left| 0\right\rangle \ \varphi _{k^{\prime
}}^{\nu }=\varepsilon _{\nu }\ \varphi _{k}^{\nu }  \label{HF}
\end{equation}
where $\left\{ ..., ...\right\}$ is the anticommutator.

It is easy to verify that (\ref{HF}) is just one of the forms of the usual
single particle HF equations, once $| 0 \rangle$ is chosen to be a
Slater determinant.

Let us now in the same way find equations which describe another form of
elementary excitations of the system such as density  vibrations. To this
purpose we define in analogy to (\ref{e}) a mean excitation energy:

\begin{equation}
E_{\mu } = 
\frac{\sum\limits_{\nu ,
k > k^\prime 
}\left\{ \left(E_{\nu
}^{N}-E_{0}^{N}\right) \left| \left\langle N,0\right| X_{kk^{\prime }}^{\mu
}a_{k}^{\dagger }a_{k^{\prime }}\left| N,\nu \right\rangle \right|
^{2}-\left( E_{\nu }^{N}-E_{0}^{N}\right) \left| \left\langle N,0\right|
Y_{kk^{\prime }}^{\mu }a_{k^{\prime }}^{\dagger }a_{k}\left| N,\nu
\right\rangle \right| ^{2}\right\} }
{\sum\limits_{\nu ,k > k^{^{\prime }}}
\left\{ \left| \left\langle N,0\right| X_{kk^{\prime }}^{\mu
}a_{k}^{\dagger }a_{k^{\prime }}\left| N,\nu \right\rangle \right|
^{2}-\left| \left\langle N,0\right| Y_{kk^{\prime }}^{\mu }a_{k^{\prime
}}^{\dagger }a_{k}\left| N,\nu \right\rangle \right| ^{2}\right\} }
\label{E}
\end{equation}

Minimization with respect to the amplitudes $X_{kk^{\prime }}^{\mu },
Y_{kk^{\prime }}^{\mu }$ leads to

\begin{equation}
\left\langle 0\right| \left[ \delta Q,\left[ H,Q_{\nu }^{\dagger }\right]
\right] \left| 0\right\rangle =E_{\nu }\ \left\langle 0\right| \left[ \delta
Q,Q_{\nu }^{\dagger }\right] \left| 0\right\rangle  \label{RPA}
\end{equation}
where

\begin{equation}
Q_{\nu }^{\dagger } =
\sum_{k > k^{\prime }}
\left( X_{kk^{\prime }}^{\nu }a_{k}^{\dagger }a_{k^{\prime}} -
Y_{kk^{\prime }}^{\nu }a_{k^{\prime
}}^{\dagger }a_{k}\right)   \label{Q}
\end{equation}
and $\delta Q$ is a variation (with respect to $X$ or $Y$) of $Q^{\dagger }$%
. Equation (\ref{RPA}) constitutes the SCRPA equations which are described
in great detail elsewhere \cite{Duk90,Duk96,Duk98a,Duk98b}. Explicitly

\begin{equation}
\left( 
\begin{array}{ll}
A & B \\ 
-B & -A
\end{array}
\right) \left( 
\begin{array}{l}
X^{\nu } \\ 
Y^{\nu }
\end{array}
\right) =E_{\nu }\ {\cal N\ }\left( 
\begin{array}{l}
X^{\nu } \\ 
Y^{\nu }
\end{array}
\right)   \label{AB}
\end{equation}
where the matrices $A$ and $B$ are double commutators coming from the left
hand side of (\ref{RPA}) and ${\cal N}$ is the norm matrix to be discussed
in the following section. They lead to a nonlinear eigenvalue problem for
the amplitudes $X$ and $Y$ which therefore have to be determined iteratively
very much like the HF eqs. (\ref{HF}). 
Equations (\ref{RPA},\ref{Q},\ref{AB}) are equivalent to
\begin{eqnarray}
\left\langle 0\right| \left[ Q_\nu,\left[ H,Q_{\nu' }^{\dagger }\right]
\right] \left| 0\right\rangle = E_{\nu } ~\delta_{\nu, \nu'} \label{QHQ}
\\
\left\langle 0\right| \left[ Q_\nu^{\dagger },\left[ H,Q_{\nu'
}^{\dagger }\right] \right] \left| 0\right\rangle = 0  \label{QHQt}
\end{eqnarray}

This form is interesting since these equations have exactly the same
structure as any mean field Hartree-Fock-Bogoliubov equations, be it for
single Fermions or Bosons or, as here, for Fermion pairs.

For a hamiltonian with two body
interactions one verifies easily that (\ref{RPA}) contains at most one and
two body density matrices. Roughly speaking the two body density matrices
can be expressed as quadratic forms of the amplitudes $X$ and $Y$ (for more
details see \cite{Duk90,Duk96,Duk98a,Duk98b}). 

An important point is to realize that (\ref{Q}) is {\bf not} restricted to
the particle-hole (ph) and (hp) subspaces as is of common use in
the nuclear literature on the subject \cite{Blai86,Row68,Rin80}. Here the
only restriction in (\ref{Q}) is that it should not contain any diagonal
(i.e. Hermitian) components. Therefore in 
$Q_{\nu }^{\dagger } =  \sum_{k \neq k^{\prime }}
 \chi_{kk^{\prime }}^{\nu }a_{k}^{\dagger }a_{k^{\prime}}$ the matrix
is not Hermitian. The single particle basis in which
(\ref{RPA},\ref{Q},\ref{AB}) shall be solved is obtained from

\begin{equation}
\left\langle 0\right| \left[ H,Q_{\nu}^{\dagger } \right]
 \left| 0\right\rangle = \left\langle 0\right| \left[ H,Q_{\nu}
\right]  \left| 0\right\rangle = 0 \label{HQ}
\end{equation}

One can show that (\ref{HQ}) is obtained from the minimization of the
SCRPA ground state energy with respect to the basis
\cite{Duk90,Duk96,Duk98b} but one also
directly realizes that (\ref{HQ}) is consistent with the equations of
motion (\ref{QHQ},\ref{QHQt}). 

The matrix $B$ contains the pair potential of the two fermion pairs whereas
the matrix $A$ contains the normal selfconsistent potential for Fermion
pairs. Qualitatively we can represent the selfconsistent equations (\ref{AB}%
) as in Fig. 1 \cite{Schu73}

Figure 1

\noindent
 where the
wiggly line stands for quantum fluctuations. Such a selfconsistent mean
field potential for density fluctuations as shown in Fig. 1 seems quite
natural, since the groundstate of an interacting Fermi system can be
considered as a gas of quantal fluctuations. The presence of fluctuations
also has a feedback on the single particle motion, an issue which we
mainly want to consider in this paper. For example, to couple back
consistently the
single particle density matrix $\rho _{kk^{\prime }}=\left\langle 0\right|
a_{k}^{\dagger }a_{k^{\prime }}\left| 0\right\rangle $ to the amplitudes $X$
and $Y$ in order to close the system of equations, has been a matter of
debate in the past\cite{Cat96}. It should be noted that, depending on the
problem at
hand, it also can happen that certain elements of the two body density
matrix can not directly be expressed via $X$ and $Y$ amplitudes. In the
Lipkin model which we will study below we will see that indeed a particular
matrix element of the two body density matrix falls into this category. We
will, however, demonstrate that once we have a method at hand that allows to
calculate the single particle density matrix we will also find a reliable
method of how to evaluate the missing two body elements.

\section{Coupling the single particle Green's function to the Self
Consistent RPA}

The eigenvalue problem (\ref{RPA}) has as usual a corresponding Green's
Function (GF) formulation. For the following it is useful to also briefly
outline this approach which, of course, is completely equivalent
to the eigenvalue problem (\ref{RPA}).

Let us therefore define the two time chronological Green's function at
zero temperature which describe density fluctuations

\begin{equation}
G_{k_{1}k_{2}k_{1}^{^{\prime }}k_{2}^{^{\prime }}}^{t-t^{\prime
}}=-i\left\langle 0\right| T\left( a_{k_{2}}^{\dagger }a_{k_{1}}\right)
_{t}\left( a_{k_{1}^{^{\prime }}}^{\dagger }a_{k_{2}^{^{\prime }}}\right)
_{t^{\prime }}\left| 0\right\rangle  \label{G1}
\end{equation}
where $T$ is the chronological operator and

\begin{equation}
O_{t}=e^{iHt}Oe^{-iHt}  \label{G2}
\end{equation}
with $H$ the full Hamiltonian operator. 
In principle in (\ref{G1}) one should take only the fluctuating operator
$a^\dagger a - \langle 0 |a^\dagger a | 0 \rangle$ but since in the
equations of motion (\ref{RPA},\ref{Q}) any c-number drops out we will
stay with the definition given in (\ref{G1}).

The Dyson equation for (\ref{G1})
corresponding to (\ref{RPA}) reads after Fourier transformation in the
approximation of the instantaneous mass operator \cite{Duk90,Duk98a}:

\begin{equation}
\omega G_{k_{1}k_{2}k_{1}^{^{\prime }}k_{2}^{^{\prime }}}^{\omega }={\cal N}%
_{k_{1}k_{2}k_{1}^{^{\prime }}k_{2}^{^{\prime
}}}+\sum_{p_{1}p_{2}k_{1}k_{2}p_{1}p_{2}}{\cal H}^{\left( 0\right) }\
G_{p_{1}p_{2}k_{1}^{^{\prime }}k_{2}^{^{\prime }}}^{\omega }  \label{G3}
\end{equation}
with

\begin{equation}
{\cal N}_{k_{1}k_{2}k_{1}^{^{\prime }}k_{2}^{^{\prime }}}=\left\langle
0\right| \left[ a_{k_{2}}^{\dagger }a_{k_{1},}a_{k_{1}^{^{\prime
}}}^{\dagger }a_{k_{2}^{^{\prime }}}\right] \left| 0\right\rangle 
\label{G4}
\end{equation}
and

\begin{equation}
{\cal H}_{k_{1}k_{2}k_{1}^{^{\prime }}k_{2}^{^{\prime }}}^{\left( 0\right)
}=\sum_{p_{1}p_{2}}\left\langle 0\right| \left[ a_{k_{2}}^{\dagger
}a_{k_{1},}\left[ H,a_{p_{1}}^{\dagger }a_{p_{2}}\right] \right] \left|
0\right\rangle \ {\cal N}_{p_{1}p_{2}k_{1}^{^{\prime }}k_{2}^{^{\prime
}}}^{-1}  \label{G5}
\end{equation}

One easily recognizes from (\ref{G3}-\ref{G5}) the equivalence with (\ref
{RPA}). Since the Eqs. (\ref{G3},\ref{G4},\ref{G5}) have been derived at
length in several preceding articles \cite{Duk90,Duk98a} we will not
represent them
here.

For the coupling with the single particle Green's function it is useful to
define a SCRPA T-matrix from (\ref{G3}) in the following way

\begin{equation}
G_{k_{1}k_{2}k_{1}^{^{\prime }}k_{2}^{^{\prime }}}^{\omega
}=G_{k_{1}k_{2}k_{1}^{\prime }k_{2}^{\prime }}^{0}\
+G_{k_{1}k_{2}p_{1}p_{2}}^{0}T_{p_{1}p_{2}p_{2}^{\prime }p_{2}^{\prime
}}^{SCRPA}\ G_{p_{1}^{\prime }p_{2}^{\prime }k_{1}^{\prime }k_{2}^{\prime
}}^{0}  \label{G6}
\end{equation}
with

\begin{equation}
G_{k_{1}k_{2}k_{1}^{\prime }k_{2}^{\prime }}^{0}=\frac{n_{k_{2}}-n_{k_{1}}}{%
\omega -\varepsilon _{1}+\varepsilon _{2}}\delta _{k_{1}k_{1}^{\prime
}}\delta _{k_{2}k_{2}^{\prime }}  \label{G60}
\end{equation}
where $n_{k}=\left\langle 0| a_{k}^{\dagger }a_{k}|0 \right\rangle $ and
$ \varepsilon _{k}=\frac{k^{2}}{2m}+\sum_{k^{^{\prime }}}\overline{v}%
_{kk^{^{\prime }}kk^{^{\prime }}}\ n_{k^{^{\prime }}}$ are the occupation
numbers and generalized single particle energies which we assumed without
loss of generality to be diagonal and $\overline{v}_{k_{1}k_{2}k_{3}k_{4}}$
is the antisymmetrised matrix element of the two body interaction. With (\ref
{G3}-\ref{G5}) the T-matrix in (\ref{G6}) is uniquely defined. Since this is
quite standard procedure we do not further elaborate on the form of the
T-matrix. A form equivalent to (\ref{G6}) is given by (we use summation
convention)

\begin{equation}
G_{k_{1}k_{2}k_{1}^{\prime }k_{2}^{\prime }}=G_{k_{1}k_{2}k_{1}^{\prime
}k_{2}^{\prime }}^{0}\
+G_{k_{1}k_{2}p_{1}p_{2}}^{0}K_{p_{1}p_{2}p_{2}^{\prime }p_{2}^{\prime
}}^{SCRPA}\ G_{p_{1}^{\prime }p_{2}^{\prime }k_{1}^{\prime }k_{2}^{\prime }}
\label{G61}
\end{equation}
with

\begin{equation}
K_{k_{1}k_{2}k_{1}^{\prime }k_{2}^{\prime }}^{SCRPA}={\cal H}%
_{k_{1}k_{2}k_{1}^{\prime }k_{2}^{\prime }}^{\left( 0\right) }-\left(
\varepsilon _{k_{1}}-\varepsilon _{k_{2}}\right) \delta _{k_{1}k_{1}^{\prime
}}\delta _{k_{2}k_{2}^{\prime }}  \label{G62}
\end{equation}

From (\ref{G61}-\ref{G62}) we also read off the equality

\begin{equation}
\sum_{k_{3}k_{4}}K_{k_{1}k_{2}k_{3}k_{4}}^{SCRPA}G_{k_{3}k_{4}k_{1}^{\prime
}k_{2}^{\prime }}=\sum_{k_{3}k_{4}}T_{k_{1}k_{2}k_{3}k_{4}}^{SCRPA}\
G_{k_{3}k_{4}k_{1}^{\prime }k_{2}^{\prime }}^{0}  \label{G63}
\end{equation}

The important point to recognize is that the mass operator of the single
particle Dyson equation

\begin{equation}
\left( \omega -\varepsilon _{k}\right) \ G_{kk^{^{\prime }}}^{\omega
}=\delta _{kk^{^{\prime }}}+\sum_{p}{\cal M}_{kp}^{\omega }\ G_{pk^{^{\prime
}}}^{\omega }  \label{G7}
\end{equation}
has a well known representation in terms of the full two body T-matrix
\cite{Duk96}. For better visibility we present the relation graphically
in figure 2.

Figure 2

At this point it has now become obvious what our interrelation of single
particle GF and SCRPA shall be: we have to replace in Fig. 2 the full
T-matrix by the approximate $T^{SCRPA}\left( \omega \right) $ defined in (%
\ref{G6}). In addition to this obvious construct there also exists a direct
and strong consistency requirement. It stems from the fact that we have now
two ways of calculating the correlation energy: the first uses the well
known relation between the single particle GF and the ground state energy
\cite{Duk98a,Jan91,Mig67}

\begin{equation}
E_{0}=-\frac{i}{2}\ \lim_{t^{\prime }-t\rightarrow 0^{+}}Tr\left( i\frac{%
\partial }{\partial t}+\varepsilon _{k}\right) \ G_{kk}^{t-t^{\prime }}\quad
\label{G9}
\end{equation}

The second expresses the correlation energy density via the two body GF (\ref
{G1}) :

\begin{equation}
E_{corr}=
{\frac i 4} \lim_{t'-t \rightarrow 0^+}
Tr\left[ \overline{v}_{k_{1}k_{2}^{\prime}k_{2}k_{1}^{\prime }}
\left( G_{k_{1}k_{2}k_{1}^{^{\prime }}k_{2}^{^{\prime}}}^{t-t' } -
G_{k_{1}k_{2}k_{1}^{^{\prime }}k_{2}^{^{\prime }}}^{(0) t-t'}  
\right) \right]   \label{G10}
\end{equation}
where again $\overline{v}_{k_{1}k_{2}^{\prime}k_{2}k_{1}^{\prime }}$ is
the antisymmetrised two-body matrix element entering in the Hamiltonian
$H$.

The requirement is now that both expressions for the correlation energy,
that is, the one deduced from (\ref{G9}) and (\ref{G10}),
agree. This is equivalent to the Hugenholtz-van Hove theorem which states
that the chemical potential $\mu $ calculated via the single particle GF
must be equal (at equilibrium ) to the energy per particle when calculated
from (\ref{G10}). It turns out that this only is achieved when expanding the
GF in (\ref{G7}) to first order in the mass operator

\begin{equation}
G_{k}=G_{k}^{0}+G_{k}^{0}\ {\cal M}_{k}^{\omega }\ G_{k}^{0}  \label{G11}
\end{equation}
with

\begin{equation}
\left( \omega -\varepsilon _{k}\right) \ G_{k}^{0}=1  \label{G12}
\end{equation}

Of course one can use the iterated solution of the Dyson equation, $i.e.$ $%
G_{k}=\left( \omega -\varepsilon _{k}-{\cal M}_{k}^{\omega }\right) ^{-1}$
but for consistency then the particle-hole propagators of the SCRPA equation
must be redefined accordingly. This has been discussed in \cite{Duk98a}
and may be elaborated in the future but for the moment we keep with the more
restrictive consistency relation (\ref{G11}) together with (\ref{G3}-\ref{G6}%
).

For space reasons we have been relatively short in this general section. We
will, however, work out in some detail the model case of the next section so
that the reader, by analogy, shall be able to reconstruct details also in
the general case quite easily.

\section{Application to the Lipkin Model}

The Hamiltonian of the Lipkin \cite{Lip65} model is given by

\begin{equation}
H=\varepsilon J_{0}-{\frac{V}{2}}(J_{+}^{2}+J_{-}^{2})  \label{lip1}
\end{equation}

\noindent with

\begin{eqnarray}
J_{0} &=&{\frac{1}{2}}\sum_{m=1}^{\Omega }(c_{1m}^{\dagger
}c_{1m}-c_{0m}^{\dagger }c_{0m})~~~,  \nonumber \\
J_{+} &=&\sum_{m=1}^{\Omega }c_{1m}^{\dagger
}c_{0m}~,~~~~~~~~~~~~~J_{-}=\sum_{m=1}^{\Omega }c_{0m}^{\dagger }c_{1m}~~~~
\label{lip2}
\end{eqnarray}

\noindent The indices 0 and 1 denote the lower and upper levels
respectively, separated by an energy $\varepsilon $, and $m$ is the angular
momentum projection in each shell with degeneracy $\Omega $.

The commutation relations between these three operators, which are the
generators of the SU(2) group, are

\begin{equation}
\lbrack J_{+},J_{-}]=2J_{0},~~~~~~~~[J_{0},J_{\pm }]=\pm J_{\pm }~~~.
\label{lip3}
\end{equation}

In the Lipkin model the number of particles is exactly that needed to
completely fill the lower shell, i.e. $N=\Omega $.

\subsection{SCRPA equations}

The SCRPA solutions are built with the operators (we stay in the normal
phase)

\begin{equation}
Q^{\dagger }={\frac{1}{\sqrt{-2 \langle J_0 \rangle }}}[X  J_{+} -
Y J_{-}]~,~~~~~
Q={\frac{1}{\sqrt{-2 \langle J_0 \rangle }}} [X J_{-}-Y J_{+}]
\label{q}
\end{equation}

\noindent acting over a correlated vacuum $|0\rangle $,which is defined by
the equation 
\begin{equation}
Q|RPA\rangle =0  \label{vac}
\end{equation}
to yield the excited state 

\begin{equation}
\left| 1\right\rangle =Q^{\dagger }|RPA\rangle ~~.
\end{equation}

The SCRPA equations (\ref{RPA}) then take the following form 
\begin{equation}
\left( 
\begin{array}{cc}
A & B \\ 
-B & -A
\end{array}
\right) \left( 
\begin{array}{c}
X \\ 
Y
\end{array}
\right) = E \left( 
\begin{array}{c}
X \\ 
Y
\end{array}
\right)   \label{SCRPA}
\end{equation}

\noindent with the matrix elements $A$ and $B$ defined by \cite{Row68,Par68} 
\begin{eqnarray}
A &=&\left\langle [J_{-},[H,J_{+\langle }]]\right\rangle /\left\langle
[J_{-},J_{+}]\right\rangle  \nonumber \\
B &=&\left\langle [J_{+},[H,J_{+}]]\right\rangle /\left\langle
[J_{-},J_{+}]\right\rangle  \label{a,b}
\end{eqnarray}
where we used $\left\langle \cdots \right\rangle $ for $\left\langle
RPA\right| \cdots \left| RPA\right\rangle $ .

The normalization of the excited state $Q^{\dagger }|RPA\rangle $ is given by

\begin{equation}
\langle QQ^{\dagger }\rangle =\langle [Q,Q^{\dagger }]\rangle
= X^{2}-Y^{2} = 1   \label{norm}
\end{equation}

With (\ref{norm}) the inversion of (\ref{q}) yields $J_{+}=\sqrt{-2
\langle J_{0}\rangle}\left(
XQ^{\dagger }+YQ\right) $ and the matrix elements of the SCRPA matrices read

\[
A_{SCRPA}=\varepsilon +2VXY~~~ 
\]

\begin{equation}
B_{SCRPA}=2V{\frac{\langle J_{0}^{2}\rangle }{\langle J_{0}\rangle }}%
+V(X^{2}+Y^{2})  \label{ab}
\end{equation}

From (\ref{ab}) we see that we face exactly the problem discussed in Sect.
3. The single particle occupation $\left\langle J_{0}\right\rangle $ and
the square $\left\langle J_{0}^{2}\right\rangle $ can not directly be
expressed
in terms of $X$ and $Y$ . Of course as is well known in the present simple
model it is possible to calculate the RPA groundstate via (\ref{vac})
explicitly \cite{Lip65,Par68} :

\begin{equation}
|RPA\rangle =\sum_{l=0}^{\Omega /2}{\frac{(\Omega -2l)!}{(\Omega /2-l)!l!}}%
~\left( {\frac{Y}{X}}\right) ^{2}~J_{+}^{2l}~|HF\rangle   \label{rpavac}
\end{equation}
and therefore also $\left\langle J_{0}\right\rangle $ and $\left\langle
J_{0}^{2}\right\rangle $ can explicitly be calculated \cite{Duk90}.
However, this is not the usual situation and in general it will be very
difficult if not impossible to solve the vacuum condition (\ref{vac}). We
have therefore to develop other methods to get access to these quantities
independently. As a word of caution we should mention again (see
Section II) that it is not
possible to include $J_{0}$ as a further component into the definition of
the RPA excitation operator $Q^{\dagger }$ (\ref{q}) since $J_{0}$ is
hermitian and it is then impossible to define the norm of the RPA excited
state.

In the next section we will therefore elaborate on the evaluation of $%
\left\langle J_{0}\right\rangle $ via the single particle GF the way we
have
outlined it in the general section 3.

For later use we also introduce here  the renormalized RPA (RRPA) matrix
elements 

\begin{equation}
A_{RRPA} = \varepsilon ,~~~~~~~
B_{RRPA} = 2 V \langle J_{0}\rangle ~~~~~~~~~~.
\end{equation}

They will be used below when we will compare the results of the SCRPA not
only to the exact solution but also to RRPA.

\subsection{SCRPA and the single particle Green's function}

\bigskip As outlined above we have to construct a mass operator for the s.
p. GF such that it yields exactly the same groundstate energy via eq.
(\ref{G9}) as
when calculated directly from the two body GF (\ref{G10}). In order to
explain the
principle we first want to exemplify the procedure with standard RPA. In
this case we have to put in eq. (\ref{ab}) $X=1$ , $Y=0$ and $\left\langle
J_{0}\right\rangle = -\frac{\Omega }{2}$, $\left\langle
J_{0}^{2}\right\rangle =\frac{\Omega ^{2}}{4}$ . Let us for example
consider
the interaction energy to RPA order

\begin{equation}
E_{pot}=-\frac{V}{2}\left( \left\langle J_{+}^{2}\right\rangle
+\left\langle J_{-}^{2}\right\rangle \right) \Rightarrow E_{pot}^{RPA}=-V\
\Omega \ X\ Y  \label{epot}
\end{equation}

Using one of the RPA equations (dropping $1/\Omega $ corrections):

\begin{equation}
V\ \Omega \ X=\left( E +\varepsilon \right) \ Y  \label{xx}
\end{equation}
and multiplying this equation with $X$ we obtain for $E_{pot\text{ }}$:

\begin{equation}
E_{pot}^{RPA}=-V\ \Omega \ \frac{X^{2}}{E+\varepsilon }\ V\ \Omega 
\label{eprpa}
\end{equation}

Expression (\ref{eprpa}) can be identified with the evaluation of the
Feynman graph shown in Figure 3

Figure 3

\noindent
where the wiggly line represents the RPA phonon with energy $E$ . The
particle-hole bubble has energy $\varepsilon _{ph}=\varepsilon $ and
together with the phonon the vertical cut has energy $E +\varepsilon $
what corresponds to the energy denominator in (\ref{eprpa}). The amplitude
of the phonon is $X^{2}$ and the two dots of the graph represents the
interaction squared. As usual we can obtain the mass operator from the
groundstate graph in cutting open the hole line. Therefore we obtain e. g.
for the GF of the upper level ( $G_{1m}^{t-t^{\prime }}=-i\left\langle
T\left( a_{1m}\left( t\right) a_{1m}^{\dagger }\left( t^{\prime }\right)
\right) \right\rangle $ ) in the approximation of eq. (\ref{G11})

\begin{equation}
G_{1m}^{\omega }=\frac{1}{\omega -\frac{\varepsilon }{2}}+\frac{1}{\omega -%
\frac{\varepsilon }{2}}\ V\ \frac{\Omega ^{2}X^{2}}{\omega +E+\frac{%
\varepsilon }{2}}\ V\ \frac{1}{\omega -\frac{\varepsilon }{2}}  \label{H14}
\end{equation}
where the mass operator has the obvious graphical representation of Fig. 4

Figure 4

Using the (exact) relation (what is just a variant of (\ref{G9})):

\begin{equation}
\frac{i}{2}\lim_{t^{\prime }-t\rightarrow 0^{+}}\left( i\frac{\partial }{%
\partial t}-\frac{\varepsilon }{2}\right) G_{1m}^{t-t^{\prime }}=-\frac{V}{2}%
\left\langle J_{+}^{2}\right\rangle  \label{H15}
\end{equation}
and inserting into the $lhs$ expression (\ref{H14}) we obtain

\begin{equation}
-i\lim_{t^{\prime }-t\rightarrow 0^{+}}\left( i\frac{\partial }{\partial t}-%
\frac{\varepsilon }{2}\right) G_{1m}^{t-t^{\prime }}=-\frac{1}{2}\left(
V\Omega \right) ^{2}\frac{X^{2}}{E+\varepsilon }  \label{H16}
\end{equation}

This is just half the potential energy (\ref{eprpa}) in the standard RPA
approach. Proceeding analogously with $%
G_{0m}$ and adding to (\ref{H16}) the corresponding expression yields the
missing
factor $2$ . This demonstrates that our construction of the mass operator in
(\ref{H14}) is consistent with the RPA groundstate energy.

Since we now have the s. p. GF at hand it is straightforward to calculate
the occupation numbers via

\begin{equation}
\left\langle a_{1m}^{\dagger }a_{1m}\right\rangle =
 -i\lim_{t^{\prime}-t\rightarrow 0^{+}} 
G_{1m}^{t-t^{\prime }}  \label{H17}
\end{equation}

Inserting into (\ref{H17}) the $rhs$ of (\ref{H14}) yields

\begin{equation}
\sum_{m}\left\langle a_{1m}^{\dagger }a_{1m}\right\rangle =\left( V\Omega
\right) ^{2}\frac{X^{2}}{\left( \varepsilon +E\right) ^{2}}=Y^{2}
\label{H18}
\end{equation}
where in the last equality we again made use of the RPA equations. It is now
easy to restore the value for $\left\langle J_{0}\right\rangle $ since $%
\sum_{m}\left\langle a_{0m}^{\dagger }a_{0m}\right\rangle =\Omega
-\sum_{m}\left\langle a_{1m}^{\dagger }a_{1m}\right\rangle $ and therefore

\begin{equation}
\left\langle J_{0}\right\rangle =-\frac{\Omega }{2}+Y^{2}  \label{H19}
\end{equation}

It is interesting to realize that (\ref{H19}) corresponds to the Holstein
Primakoff boson expansion of $\left\langle J_{0}\right\rangle $
\cite{Rin80}, a result which of course is consistent with RPA theory.

Let us now repeat the same procedure but with SCRPA. Using the SCRPA
equations, in analogy to the steps above, we can write for $E_{pot}$ :

\begin{equation}
E_{pot}^{SCRPA}=-2\left\langle J_{0}\right\rangle V\frac{\widetilde{A\ }%
XY+B\ X^{2}}{E+\varepsilon }  \label{H20}
\end{equation}
where $\widetilde{A}=A-\varepsilon $ and $A,B$ are determined in (\ref{ab}).
Again in cutting open the hole line we now find in analogy with (\ref{H14})
for the mass operator according to (\ref{G11})

\begin{equation}
G_{1m}^{\omega }=\frac{1}{\omega -\frac{\varepsilon }{2}}-\frac{1}{\omega -%
\frac{\varepsilon }{2}}\ V\frac{\left\langle 0\right| J_{-}\left|
1\right\rangle }{\omega +E+\frac{\varepsilon }{2}}\left[ \left\langle
0\right| J_{-}\left| 1\right\rangle \widetilde{A}+\left\langle 1\right|
J_{+}\left| 0\right\rangle B\right] \ \frac{1}{\omega -\frac{\varepsilon
}{2}%
}  \label{H21}
\end{equation}
where $\left| 1\right\rangle $ again is the excited state $Q^{\dagger
}\left| 0\right\rangle $ . In the RPA limit we obtain (\ref{H14}). We
immediately check that indeed we get back from $G_{1m}$ ( and $G_{0m}$ ) the
correct expression (\ref{H20}) for $E_{pot}^{SCRPA}$ inserting (\ref{H21})
into the $lhs$ of (\ref{H15}) ( and similar for $G_{0m}$ ).

Since we now have a consistent SCRPA expression for the single particle GF
at hand we proceed, as this was our goal, to the calculation of $%
\left\langle J_{0}\right\rangle $. Inserting (\ref{H21}) into the $rhs$ of
(%
\ref{H17}) one directly obtains

\begin{eqnarray}
\left\langle J_{0}\right\rangle = &-\frac{\frac{\Omega }{2}}{1-2\left[ 
\widetilde{A\ }XY+B\ X^{2}\right] \frac{V}{\left( \varepsilon
+E \right) ^{2}} }  \label{H22}\\
= &-\frac{\frac{\Omega }{2}}{1 + 2 XY \frac{V}{\left(
\varepsilon +E \right) } } \nonumber
\end{eqnarray}

Of course this is still an implicit equation for $\left\langle
J_{0}\right\rangle $, since the SCRPA eigenvalue $E$ depends on it. 
 Before proceeding it is
interesting to study several limits of (\ref{H22}). Of course for the
interaction going to zero we recover the free gas limit $\left\langle
J_{0}\right\rangle $ $=-\frac{\Omega }{2}$ . We already checked that (\ref
{H21}) goes over into the RPA limit (\ref{H14}) when $\widetilde{A}$, $B$
and the transition amplitudes are replaced by their RPA expressions.
Therefore we also recover the boson expansion result.

One should note that in order to obtain the correct RPA result one must
not make the mistake to
go over to the RPA limit, i. e. $X=1$ , $Y=0$ , $\left\langle
J_{0}^{2}\right\rangle $ $=\left\langle J_{0}\right\rangle
^{2}=\frac{\Omega
^{2}}{4}$ only in (\ref{H22}) because to get (\ref{H22}) already the
assumption has been used that $\left\langle J_{0}\right\rangle $ $\neq -%
\frac{\Omega }{2}$ on the $rhs$ of (\ref{H21}) what would not be consistent
with the RPA groundstate energy then. Now if we nevertheless take the RPA
limit, using directly (\ref{H22}), one obtains

\begin{equation}
\left\langle J_{0}\right\rangle =-\frac{\frac{\Omega
}{2}}{1+\frac{2}{\Omega 
}Y^{2}}  \label{H23}~~~~~~~.
\end{equation}

This result is interesting because it is precisely the lowest order result
which one obtains with the number operator method \cite{Cat96,Cat94}. In
the light of our theory this formula (\ref{H23}) seems to be inconsistent
because if on the $rhs$ of (\ref{H21}) one keeps $\left\langle
J_{0}\right\rangle $ $\neq -\frac{\Omega }{2}$ , there is no reason to
drop
all the other terms going beyond standard RPA. So in this light the pure
lowest order boson result (\ref{H19}) seems to be more consistent than the
partially resummed series (\ref{H23}). We will see later that this is indeed
confirmed by numerical results.

\subsection{Determination of $\left\langle J_{0}^{2}\right\rangle $}

In principle we are still short of the expectation value of the square of
the occupation number.
Eventually we could try to establish an analogous expression to what has
been found for $\left\langle J_{0}\right\rangle $ (\ref{H21}). However, at
least in the present model the factorization relation

\begin{equation}
\left\langle J_{0}^{2}\right\rangle \cong \left\langle J_{0}\right\rangle
^{2}  \label{H24}
\end{equation}
seems to be extremely well fulfilled for the whole range of the interaction
strength considered (see next section). Of course this may be a
particularity of the model but we suppose that, as long as the operator $%
J_{0}$ or analogous operators in other problems are sufficiently
collective,
equation (\ref{H24}) should work quite reasonably. In order to check this we
present the ratio $r = - \sqrt{\left\langle J_{0}^{2}\right\rangle}
/\left\langle J_{0}\right\rangle $ 
for the fixed interaction strength $%
\chi = V\left( \Omega -1\right) /\varepsilon = 1.$ (i. e. at the meanfield
transition point where fluctuations are expected to be maximal) as a
function of $\Omega $ in Figure 5. The exact results are represented by
full squares, those obtained using the exact RPA vacuum (\ref{rpavac}) by
a full line (SCRPA) and a dotted line (RRPA).

Figure 5

Only for $\Omega $ values lower than 4 one can see a significant
deviation from unity. So definitely s-wave shells are difficult 
candidates.
On the other hand should there be no degeneracy at all like in a rotating
nucleus or in an electron system in a magnetic field there is no need to
know the occupation number square since we have anyway

\begin{equation}
\left\langle a_{k}^{\dagger }a_{k}a_{k}^{\dagger }a_{k}\right\rangle
=\left\langle a_{k}^{\dagger }a_{k}\right\rangle  \label{H25}
\end{equation}

So unless there is appearance of two fold degenerate levels in a problem one
is probably well off with the factorization (\ref{H24}). In the former case
a perturbative expansion of square operators in terms of linear operators as
proposed in \cite{Duk90} using RPA excited states as intermediate
states should adequately improve on (\ref{H24}) which represents the zero
order approximation. This approximation is based on expanding the
expectation value of any two body operator by inserting a complete set of
RPA states. Specifically for the Lipkin model we have \cite{Duk90}

\begin{equation}
\left\langle J_{0}^{2}\right\rangle =\sum_{l}\frac{\left| \left\langle
J_{0}Q^{\dagger 2l}\right\rangle \right| }{\left\langle Q^{2l}Q^{\dagger
2l}\right\rangle }  \label{H251}
\end{equation}

Truncating to first order and evaluating the expectation values using the
vacuum condition we finally arrive to

\begin{equation}
\left\langle J_{0}^{2}\right\rangle =\left\langle J_{0}\right\rangle
^{2}+%
\frac{4XY\left\langle J_{0}\right\rangle ^{2}}{2\left\langle
J_{0}^{2}\right\rangle +\left( X^{2}+Y^{2}\right) \left\langle
J_{0}\right\rangle }  \label{H252}
\end{equation}
a relation which expresses $\left\langle J_{0}^{2}\right\rangle $ in terms
of $\left\langle J_{0}\right\rangle .$

Let us next study the numerical results as they follow from our SCRPA
theory described above.

\section{Numerical results}

In this section we mostly will present results for $\Omega = 14$. We will
begin
in first place to investigate the quality of the results for the correlation
part of the groundstate energy, i. e. the correlation energy

\begin{equation}
E_{corr}=\left\langle H\right\rangle -\varepsilon \frac{N}{2}  \label{I1}
\end{equation}
with

\begin{equation}
\left\langle H\right\rangle =\left\langle J_{0}\right\rangle \left[
\varepsilon -V\ X\ Y\right]   \label{I2}
\end{equation}

Figure 6

We show $E_{corr}$ as a function of $\chi   = V\left( \Omega
-1\right) /\varepsilon $ in Fig. 6 for the  RPA (dashed line) ,
RRPA (small dots), SCRPA (full line) and the exact solution (full
squares). It is a very
well known fact that RPA due to the quasiboson approximation, i. e. the
violation of the Pauli principle, overestimates in general quite strongly
the correlations and in fact overbinds in the groundstate energy. This the
more, the closer one comes to the phase transition point where RPA
collapses. This strong overbinding of the RPA was for example also found in
a recent calculation \cite{Gue96} of the electronic binding
energy of a metallic cluster. When compared with the exact results SCRPA
performs extremely well for $E_{corr}$ up to and even beyond the mean
field phase transition point $\chi = 1$ whereas RRPA starts to deviate
strongly from the exact result at $\chi\cong 1$ . 

Since it is not possible to distinguish in Fig. 6 that the SCRPA values
of $E_{corr}$ stays consistently above the exact ones, we also present the
results in Table 1.

Table 1

Another interesting quantity is the excitation energy. We show $%
E$ as a function of $\chi $ in Fig. 7. A similar scenario as in the
previous figure prevails: SCRPA yields
by far the best agreement with the exact results though the differences for $%
\chi \gtrsim 1$ are now more pronounced. It is also true
here
 that the SCRPA
excitation energy stays consistently above the exact results as can be
seen from Table 2. 

Figure 7

One
could
 conclude from that that the SCRPA also leads to an upper bound for
the excitation energy. This conjecture may be backed from the fact that we
actually derived in Section 2 the SCRPA equations from a minimization
within respect an average excitation energy. However, before drawing any
definite conclusion in this respect, a more general model with more levels
must be studied. 

Table 2

Let us now come to the investigation of the quality of the different
expressions for $\left\langle J_{0}\right\rangle $. There are essentially
three: the one which we prefer on theoretical grounds is the one from the
Green's function approach (\ref{H22}), since it is the only one which
fulfills a strong consistency relation with SCRPA equations (i. e. the
Hugenhotlz-van Hove theorem). The second is the quasiboson approximation (%
\ref{H19}) which represents the lowest order correction in $1/\Omega $ to
the free gas results. The third comes from the so-called number operator
expression (\ref{H23}) which has recently become very popular in the nuclear
physics literature \cite{Cat96,Cat94,Toi95}. We have shown that it is as
well obtainable from the GF approach in operating additional approximations to
(\ref{H22}), and that those approximations are not consistent among
them. 

Figure 8

In Fig. 8 we show the quantity $\Omega /2+\left\langle
J_{0}\right\rangle $ as a function of $\chi $ for the three approximations
to $\left\langle J_{0}\right\rangle $ when used in the SCRPA equations
(of course only the one corresponding to GF method corresponds to our
definition proper of SCRPA). In addition we show in Fig. 8 also the exact
result (full squares). The solution of the GF method (and therefore the
SCRPA proper) is shown
by the full line. The quasiboson approximation is shown by the broken line
and the number operator method by the dotted line. Not unexpectedly the GF
results are closest to the exact ones. Somewhat a
surprise is that the number operator method works no better than the
quasiboson approximation. However in the light of our discussion in
section 4 where we
argue that one passes from the GF expression (\ref{H22}) in an essentially
uncontrolled way to the number operator expression (\ref{H23}) this outcome
may seem less astonishing.

We should also say that the injection of $\left\langle J_{0}\right\rangle
$
and $\left\langle J_{0}^{2}\right\rangle $ as expressed with the RPA
groundstate wavefunction (\ref{rpavac}) into the SCRPA equation
still improves the results in Fig. 8 with respect to GF. However, we do
not show this result in order not to overload the figure and because it
corresponds to a situation which in general is not realizable.

In Figure 9  $\Omega /2+\left\langle J_{0}\right\rangle $ is shown not as
a function of $\chi$ for fixed $\Omega $ but for
fixed $\chi = 1.$ as a function of $\Omega $ . 

Figure 9

Again we see that $\Omega =2$ appears as the worst case. It is, however,
interesting to see that for this case the differences between the various
approximations are also largely enhanced without, however, inverting their
respective order.

One last interesting quantity is the ratio $Y/X$ as a function of $\chi $,
shown in Fig. 10.
It is well known that this ratio goes to 1 when approaching the phase
transition point in RPA (as seen in th broken line) while the value of
$X$ and $Y$ tend to $\infty $ individually. This then makes any RPA result
close to a phase transition
meaningless. On the other hand in SCRPA this ratio still stays of the order $%
1/2$ around the transition point and also $X$ and $Y$ remain within very
reasonable limits ($X=1.156, Y=0.580$ at $\chi = 1.$)

Figure 10

A word of caution is worth here. While the energetics and the occupation
numbers obtained with the SCRPA are very close to the exact ones, the
wave functions around and beyond $\chi =1$ (the value at which
standard RPA collapses), being far better than those obtained with RPA
or RRPA, can nontheless have an overlap with the exact wave function of
less than 50\% \cite{Hir99}. In this case the SCRPA 
must be extended to the deformed basis \cite{Duk90}.

\section{Conclusions}

In this work we addressed the question of how to close the SCRPA equations
in a consistent way and, in particular, of how to calculate single
particle quantities such as occupation numbers in this formalism. We showed
in detail how to couple back SCRPA into the single particle propagator 
consistently. The consistency criterion was based on the fulfillment of
the Hugenholtz-van Hove theorem which states that the chemical potential
obtained from the single particle propagator must be equal (at
equilibrium) to the energy per particle when directly calculated via the
correlation
function. For some problems (for instance in such schematic models as
considered here) there may also be correlation functions which involve the
expectation value of the square of the occupation number operator, which
fall out of the SCRPA space. We, however, showed that in general it seems to
be an excellent approximation to replace the expectation values of these
operators squared by the product of expectation values of the individual
operators. Only for the very special case of $\Omega =2$ we found that
some
caution has to prevail, though a perturbative expansion has been already
proposed (Eq. (\ref{H251})) to improve this approximation when needed.

Concerning the numerical results we found that SCRPA yields for this model
case excellent results (besides $\Omega =2$, see above). For instance we
found that groundstate as well as excited energies are always close but
consistently above the exact values. We also calculated the occupation
numbers from the
proposed form of the single particle propagator and found that they are
closest to the exact values in comparison with other proposed approximate
forms for the occupation numbers. Somewhat as a surprise comes the fact that
the so called number operator method yields results not better
than the quasiboson approximation. We give reasons which may back that this
is in fact a generic feature. One should say, however, that the numerical
differences for the occupation numbers using the different methods are, at
least for the model considered, not very pronounced.

We also should mention that it is not very difficult to obtain good
results for the Lipkin model in incorporating groundstate correlations in
one way or the other. However, at comparable numerical complexity, the
SCRPA equations do at least equally well, if not better, than any other
theory on the market. In this respect we refer the reader to our earlier
study of ref. \cite{Duk90}. 
A more severe test would be to apply the present SCRPA
scheme to other more realistic models like for example the multilevel
pairing model for which, in the superfluid phase, the number operator
approximation is not anymore valid. Such studies shall be presented in
future work. 

\smallskip

{\large Acknowledgments}

This work was supported in part by Conacyt (M\'{e}xico) and by the DIGICYT
(Spain) under contract No PB95/0123. J.G.H. thanks the warm hospitality
received during his visit at the IEM, CSIC, Madrid, where part of this work
was done.

\newpage
\centerline{\large Table Captions}

\bigskip
Table 1: Correlation energy $E_{corr}$ as a function of the interaction
strength $\chi$.

\bigskip
Table 2: Excitation energy $E$ as a function of the interaction strength
$\chi$.

\newpage

\centerline{Table 1}

\bigskip
\begin{tabular}{c||c|c|c|c}
 $~~\chi ~~~$   & ~~~RPA~~~& ~~~RRPA~~~ & ~~~SCRPA~~~&~~~exact~~~ \\
\hline \hline
 0.00  &   0.00000 &  0.00000  &   0.00000  &  0.00000  \\
 0.05  &  -0.00072 & -0.00067  &  -0.00067 & -0.00067 \\
 0.10  &  -0.00289 & -0.00270  &  -0.00270 & -0.00270 \\
 0.15  &  -0.00653 & -0.00608  &  -0.00608 & -0.00608 \\
 0.20  &  -0.01167 & -0.01083  &  -0.01083 & -0.01084 \\
 0.25  &  -0.01836 & -0.01698  &  -0.01698 & -0.01700 \\
 0.30  &  -0.02666 & -0.02455  &  -0.02456 & -0.02458 \\
 0.35  &  -0.03665 & -0.03358  &  -0.03359 & -0.03364 \\
 0.40  &  -0.04846 & -0.04409  &  -0.04411 & -0.04420 \\
 0.45  &  -0.06221 & -0.05612  &  -0.05618 & -0.05634 \\
 0.50  &  -0.07809 & -0.06972  &  -0.06985 & -0.07011 \\
 0.55  &  -0.09635 & -0.08490  &  -0.08517 & -0.08560 \\
 0.60  &  -0.11731 & -0.10166  &  -0.10220 & -0.10288 \\
 0.65  &  -0.14142 & -0.11996  &  -0.12101 & -0.12206 \\
 0.70  &  -0.16932 & -0.13963  &  -0.14165 & -0.14325 \\
 0.75  &  -0.20199 & -0.16029  &  -0.16418 & -0.16660 \\
 0.80  &  -0.24103 & -0.18117  &  -0.18864 & -0.19224 \\
 0.85  &  -0.28936 & -0.20064  &  -0.21506 & -0.22035 \\
 0.90  &  -0.35353 & -0.21552  &  -0.24344 & -0.25111 \\
 0.95  &  -0.45504 & -0.21994  &  -0.27375 & -0.28474 \\
 1.00  & 	      & -0.20449  &  -0.30596 & -0.32145 \\
 1.05  &   	      & -0.15752  &  -0.34001 & -0.36151 \\
 1.10  &   	      & -0.07081  &  -0.37584 & -0.40517 \\
 1.15  &           &  0.05370  &  -0.41340 & -0.45271 \\
 1.20  &           &  0.20349  &  -0.45263 & -0.50440 \\
 1.25  &           &  0.36207  &  -0.49349 & -0.56051 \\
 1.30  &           &  0.51542  &  -0.53594 & -0.62129
\end{tabular}
\newpage

\centerline{Table 2}

\bigskip

\begin{tabular}{c||c|c|c|c}
 $~~\chi ~~~$   & ~~~RPA~~~& ~~~RRPA~~~ & ~~~SCRPA~~~&~~~exact~~~ \\
\hline \hline
 0.00  &   1.00000 &  1.00000  &   1.00000 &  1.00000 \\
 0.05  &   0.99875 &  0.99875  &   0.99894 &  0.99894 \\
 0.10  &   0.99499 &  0.99499  &   0.99577 &  0.99577 \\
 0.15  &   0.98869 &  0.98870  &   0.99048 &  0.99048 \\
 0.20  &   0.97980 &  0.97986  &   0.98308 &  0.98308 \\
 0.25  &   0.96825 &  0.96840  &   0.97359 &  0.97356 \\
 0.30  &   0.95394 &  0.95426  &   0.96202 &  0.96194 \\
 0.35  &   0.93675 &  0.93737  &   0.94840 &  0.94821 \\
 0.40  &   0.91652 &  0.91762  &   0.93290 &  0.93240 \\
 0.45  &   0.89303 &  0.89491  &   0.91539 &  0.91450 \\
 0.50  &   0.86603 &  0.86908  &   0.89605 &  0.89455 \\
 0.55  &   0.83516 &  0.83998  &   0.87500 &  0.87258 \\
 0.60  &   0.80000 &  0.80744  &   0.85242 &  0.84862 \\
 0.65  &   0.75993 &  0.77126  &   0.82852 &  0.82275 \\
 0.70  &   0.71414 &  0.73126  &   0.80358 &  0.79503 \\
 0.75  &   0.66144 &  0.68728  &   0.77795 &  0.76555 \\
 0.80  &   0.60000 &  0.63925  &   0.75200 &  0.73444 \\
 0.85  &   0.52678 &  0.58729  &   0.72616 &  0.70184 \\
 0.90  &   0.43589 &  0.53194  &   0.70088 &  0.66793 \\
 0.95  &   0.31225 &  0.47437  &   0.67658 &  0.63290 \\
 1.00  &           &  0.41663  &   0.65362 &  0.59701 \\
 1.05  &           &  0.36150  &   0.63226 &  0.56050 \\
 1.10  &           &  0.31175  &   0.61269 &  0.52369 \\
 1.15  &           &  0.26907  &   0.59499 &  0.48690 \\
 1.20  &           &  0.23372  &   0.57914 &  0.45046 \\
 1.25  &           &  0.20494  &   0.56507 &  0.41472 \\
 1.30  &           &  0.18157  &   0.55267 &  0.38001
\end{tabular}

\newpage \centerline{\large Figure Captions}

\bigskip

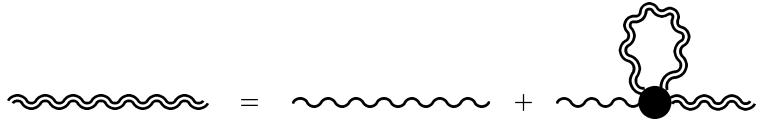
Figure 1:  Selfconsistent mean field potential for quantum fluctuations.

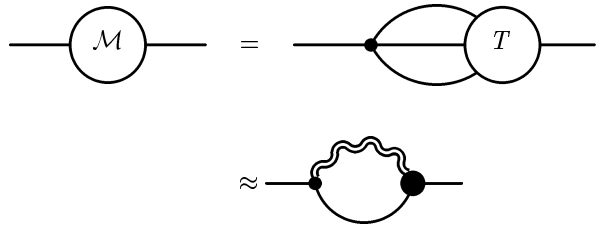
Figure 2: The mass operator of the single particle Dyson equation
represented in terms of the full two body T-matrix.

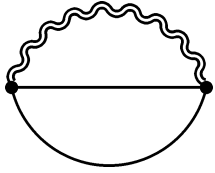
Figure 3: Feynman graph representing $E_{pot}^{RPA}$.

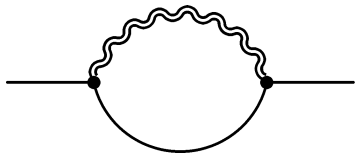
Figure 4: Groundstate  graph for the mass operator $G_{1m}^{\omega }$.

Figre 5: The ratio $r=-\sqrt{\left\langle
J_{0}^{2}\right\rangle } /\left\langle J_{0}\right\rangle $ for the fixed
interaction strength $\chi = V\left( \Omega -1\right) /\varepsilon =1$ as
a function of $\Omega$.

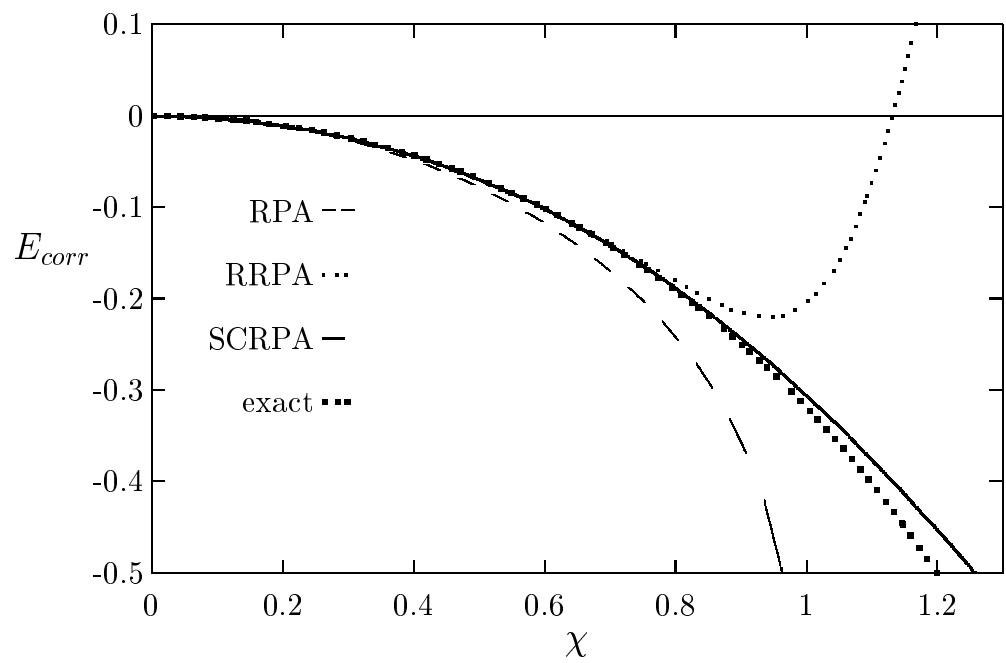
Figure 6: Correlation energy $E_{corr}$ vs. the interaction strength
$\chi$, using the exact solutions (full squares), the
RPA (dashed  line), the RRPA (small dots) and the SCRPA (full line).

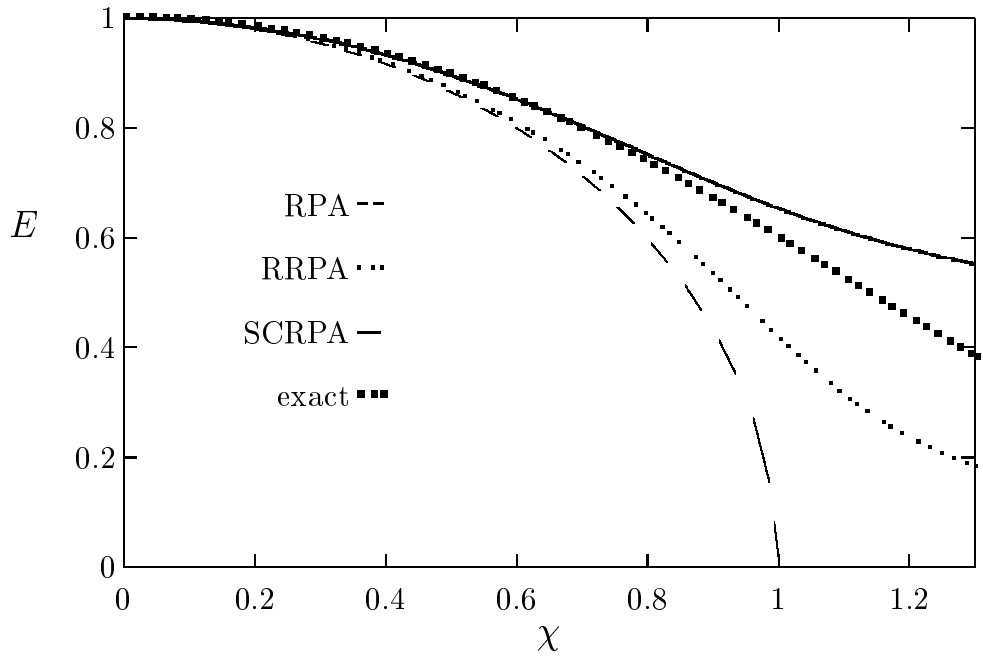
Figure 7: Excitation energy $E$ vs. the interaction strength
$\chi$, with the same convention of Fig. 6.

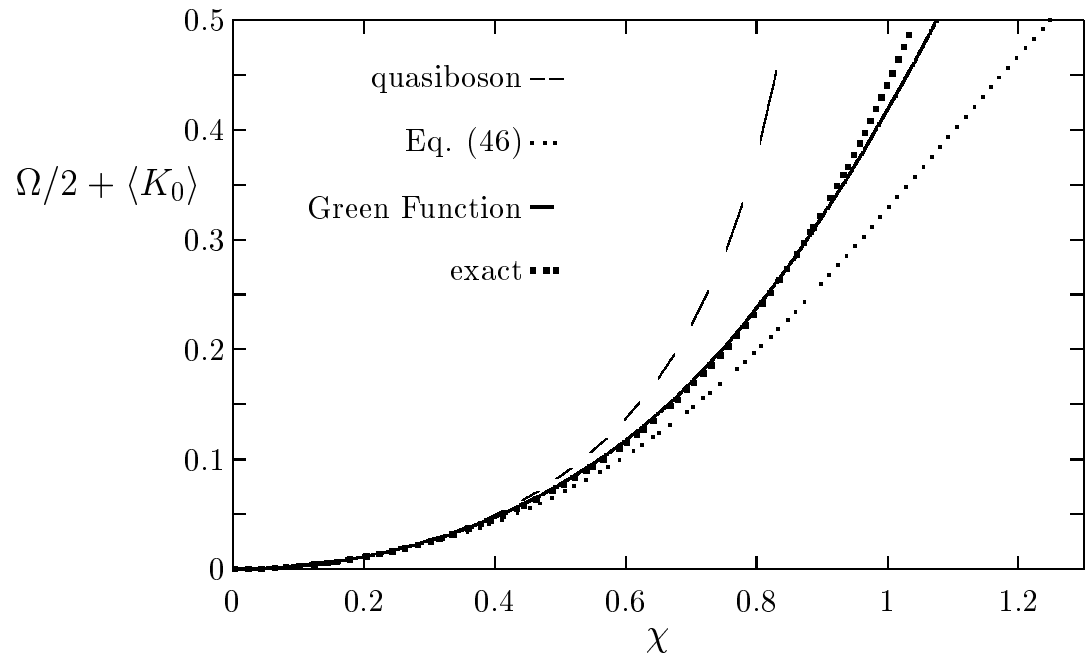
Figure 8: Occupation numbers $\Omega /2+\left\langle
J_{0}\right\rangle $ as a function of $\chi $. 

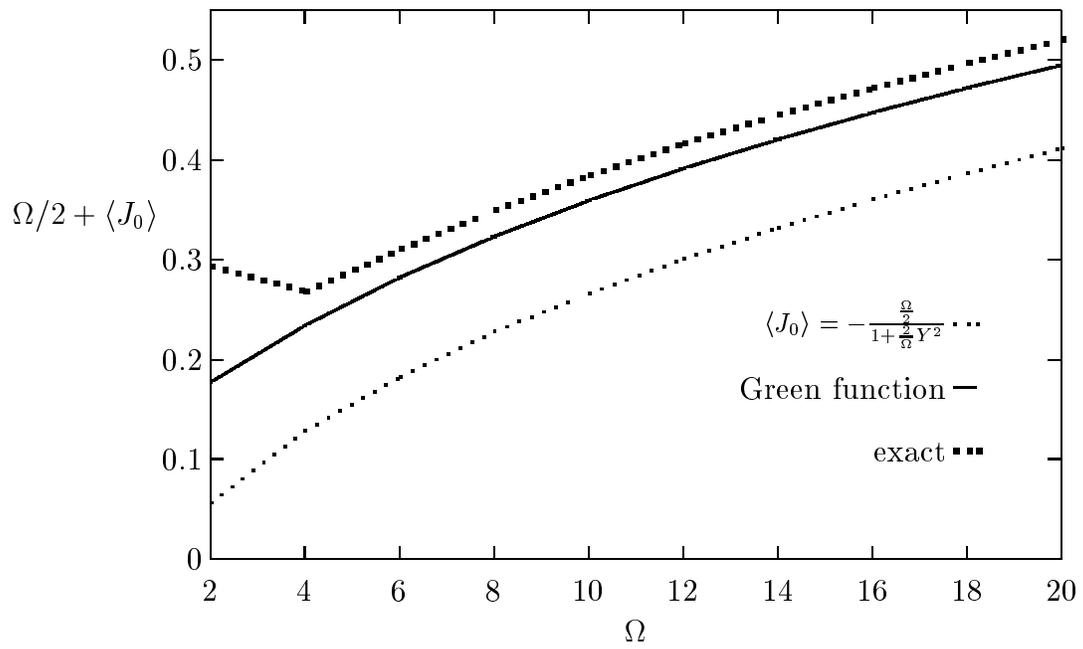
Figure 9: Occupation numbers $\Omega /2+\left\langle J_{0}\right\rangle $
as a function of $\Omega $ for fixed $\chi = 1$.

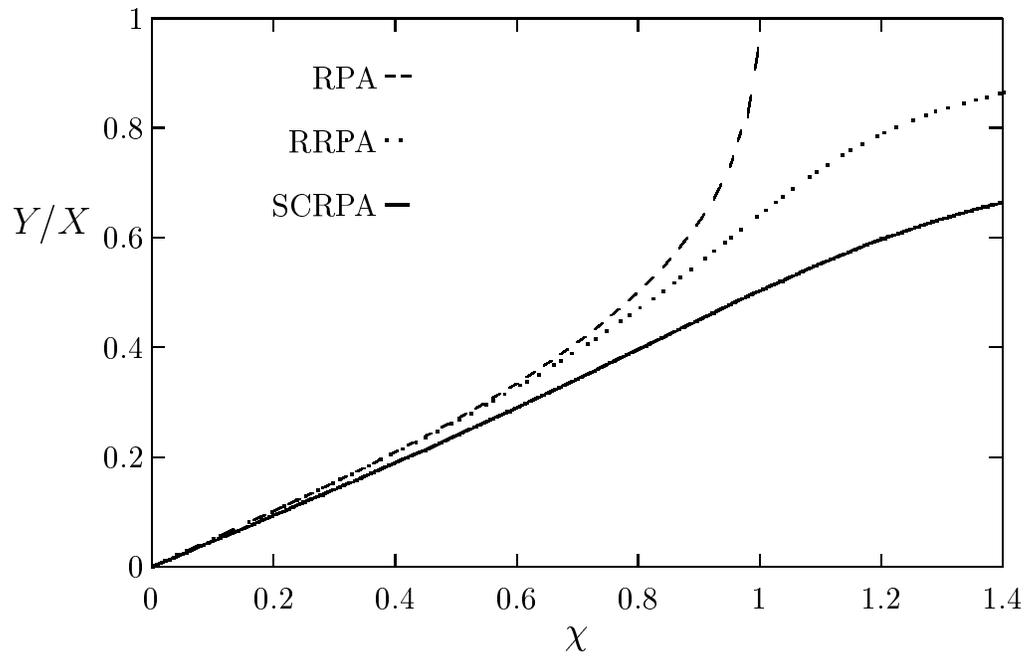
Figure 10: RPA components of the wave function $Y/X$ as function of the
interaction strength. The lines follow the same convention as in Fig. 6.

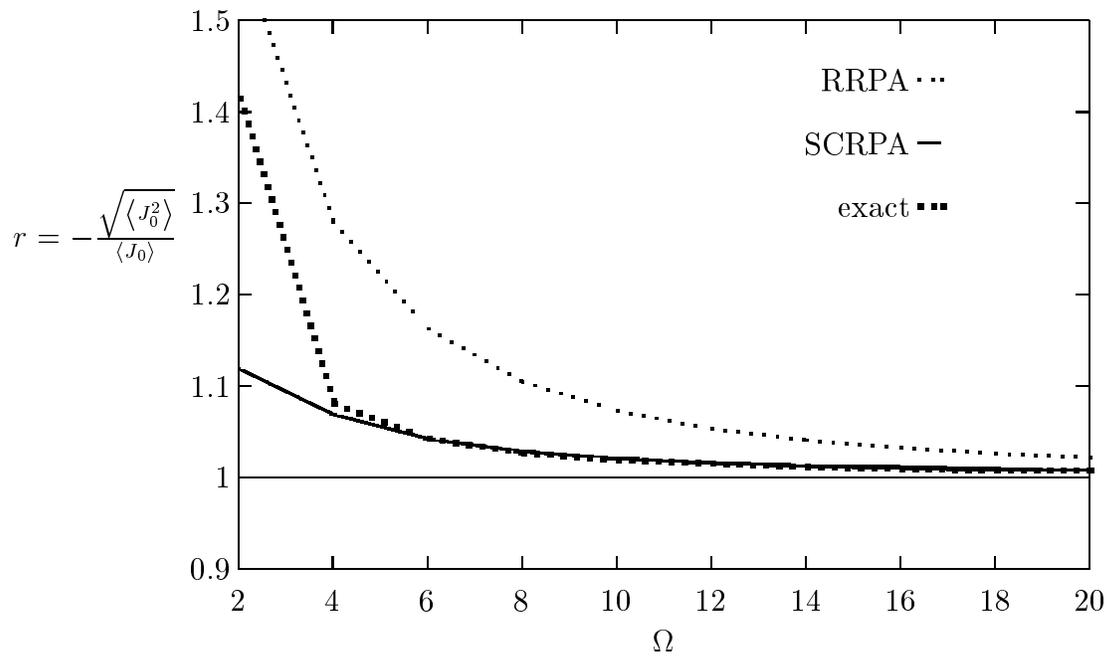
\end{document}